\documentclass[11pt, draftclsnofoot, onecolumn]{IEEEtran}

\usepackage{setspace}
\usepackage{cite,graphicx,amsmath,amssymb}
\usepackage{subfigure}
\usepackage{citesort}
\usepackage{fancyhdr}
\usepackage{mdwmath}
\usepackage{mdwtab}
\usepackage{balance}
\usepackage{xcolor}
\usepackage{bm}
\usepackage{amsthm}
\usepackage{threeparttable}
\usepackage{algorithm}
\usepackage{algorithmic}
\usepackage{multirow}
\usepackage{flafter}
\usepackage{makecell}
\usepackage{diagbox}

\providecommand{\url}[1]{#1}
\allowdisplaybreaks
\begin{document}
\title{STAR: Simultaneous Transmission And Reflection for 360$^{\circ}$ Coverage by Intelligent Surfaces}

\author{
Yuanwei~Liu, Xidong~Mu, Jiaqi~Xu, Robert~Schober, Yang~Hao, H.~Vincent~Poor, and Lajos~Hanzo

\thanks{Y. Liu, J. Xu, and Y. Hao are with the School of Electronic Engineering and Computer Science, Queen Mary University of London, London E1 4NS, UK, (email: \{yuanwei.liu, jiaqi.xu, y.hao\}@qmul.ac.uk).}
\thanks{X. Mu is with School of Artificial Intelligence, Beijing University of Posts and Telecommunications, Beijing, 100876, China (email: muxidong@bupt.edu.cn).}
\thanks{R. Schober is with the Institute for Digital Communications, Friedrich-Alexander-University Erlangen-N{\"u}rnberg (FAU), Germany (e-mail: robert.schober@fau.de).}
\thanks{H. V. Poor is with the Department of Electrical Engineering, Princeton University, Princeton, NJ 08544 USA (e-mail: poor@princeton.edu).}
\thanks{L. Hanzo is with the school of Electronics and Computer Science, University of Southampton, Southampton SO17 1BJ, U.K. (e-mail: lh@ecs.soton.ac.uk).}
}

\maketitle
\begin{abstract}
A novel simultaneously transmitting and reflecting (STAR) system design relying on reconfigurable intelligent surfaces (RISs) is conceived. First, an existing prototype is reviewed and the potential benefits of STAR-RISs are discussed. Then, the key differences between conventional reflecting-only RISs and STAR-RISs are identified from the perspectives of hardware design, physics principles, and communication system design. Furthermore, the basic signal model of STAR-RISs is introduced, and three practical protocols are proposed for their operation, namely energy splitting, mode switching, and time switching. Based on the proposed protocols, a range of promising application scenarios are put forward for integrating STAR-RISs into next-generation wireless networks. By considering the downlink of a typical RIS-aided multiple-input single-output (MISO) system, numerical case studies are provided for revealing the superiority of STAR-RISs over other baselines, when employing the proposed protocols. Finally, several open research problems are discussed.
\end{abstract}

\section{Introduction}
With the rapid development of metasurfaces and the corresponding fabrication technologies, reconfigurable intelligent surfaces (RISs) and their diverse variants~\cite{Renzo2020JSAC,di2019smart,Huang_Holographic,Wu_MAG} have emerged as promising techniques for sixth-generation (6G) wireless networks. Generally speaking, RISs are two-dimensional (2D) structures and are comprised of a large number of low-cost reconfigurable elements. By employing a smart controller (e.g., a field-programmable gate array (FPGA)) attached to the RIS, both the phase and even the amplitude of these reconfigurable elements can be beneficially controlled, thus reconfiguring the propagation of the incident wireless signals and realizing a ``Smart Radio Environment (SRE)''~\cite{Renzo2020JSAC,di2019smart}. Since no radio frequency (RF) chains are required, RISs are more economical and environmentally friendly than the family of conventional multi-antenna and relaying technologies. Given these beneficial RIS characteristics, extensive industrial and academic research efforts have been devoted to the investigation of RISs, including but not limited to the design of energy efficient communication~\cite{Huang_EE}, the mitigation of blockages in millimeter wave (mmWave)/terahertz (THz) communications~\cite{Akyildiz}, and their artificial intelligence (AI) aided implementation~\cite{Huang_AI}.\\
\indent Although some recent studies have considered both transmissive and reflective metasurfaces for wireless communications (as will be discussed below), existing contributions mainly focus on RISs whose only function is to reflect an incident signal, hence both the source and the destination have to be at the same side of the RISs~\cite{Huang_EE,Akyildiz,Huang_AI}, i.e., within the same \emph{half-space} of the SRE. This topological constraint limits the flexibility of employing conventional RISs, and to address this issue, we propose the concept of simultaneous transmitting and reflecting RISs (STAR-RISs)\footnote{STAR-RISs were termed as simultaneously refracting and reflecting RISs (SRAR-RISs) in our previous work~\cite{SRAR}. Since ``transmitting'' and ``refracting'' have the same meaning from a physical perspective, we use ``transmitting'' in the rest of the article for convenience.}, where the incident wireless signals can be reflected within the half-space of the SRE at the same side of the RIS, but they can also be transmitted into the other side of the RIS. As a result, a \emph{full-space} SRE can be created by STAR-RISs.
\begin{figure*}[!t]
\centering
\subfigure[]{\label{prototype}
\includegraphics[width= 2.8in]{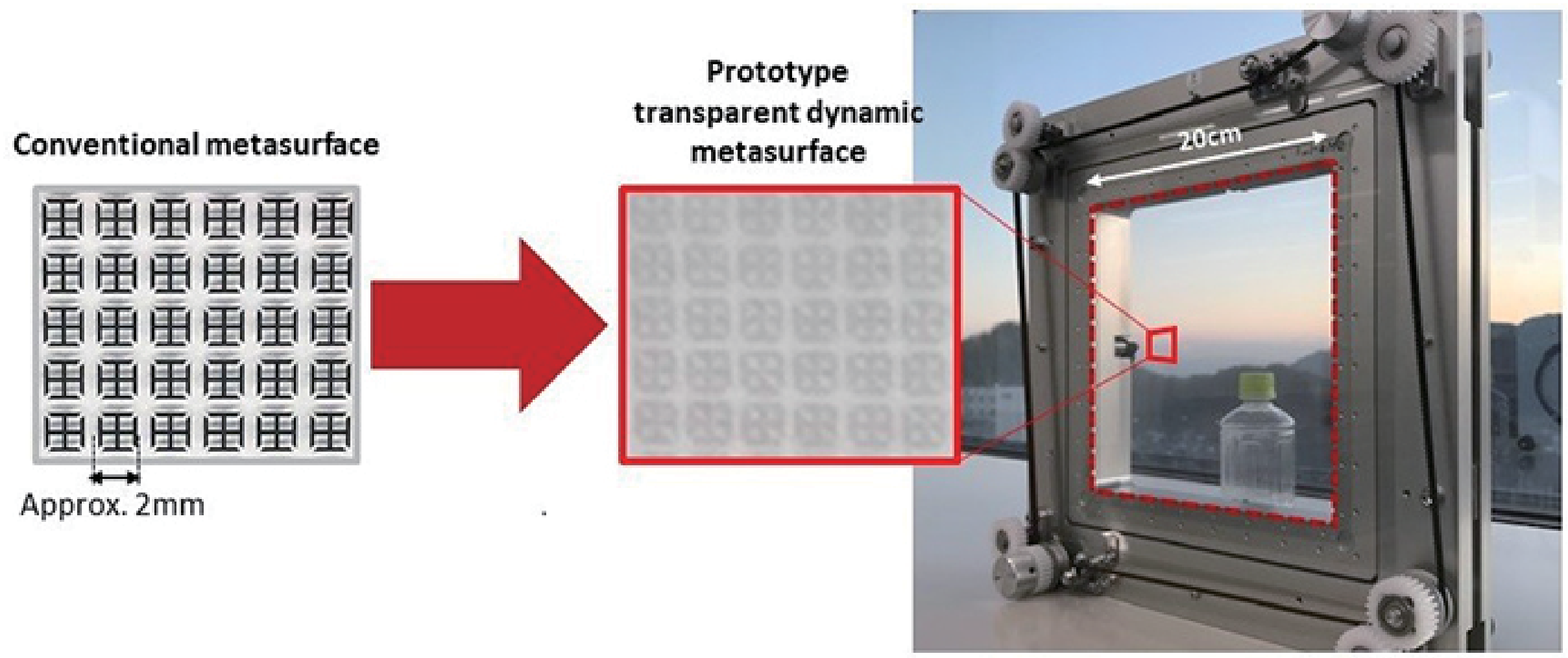}}
\subfigure[]{\label{FR}
\includegraphics[width= 1.1in]{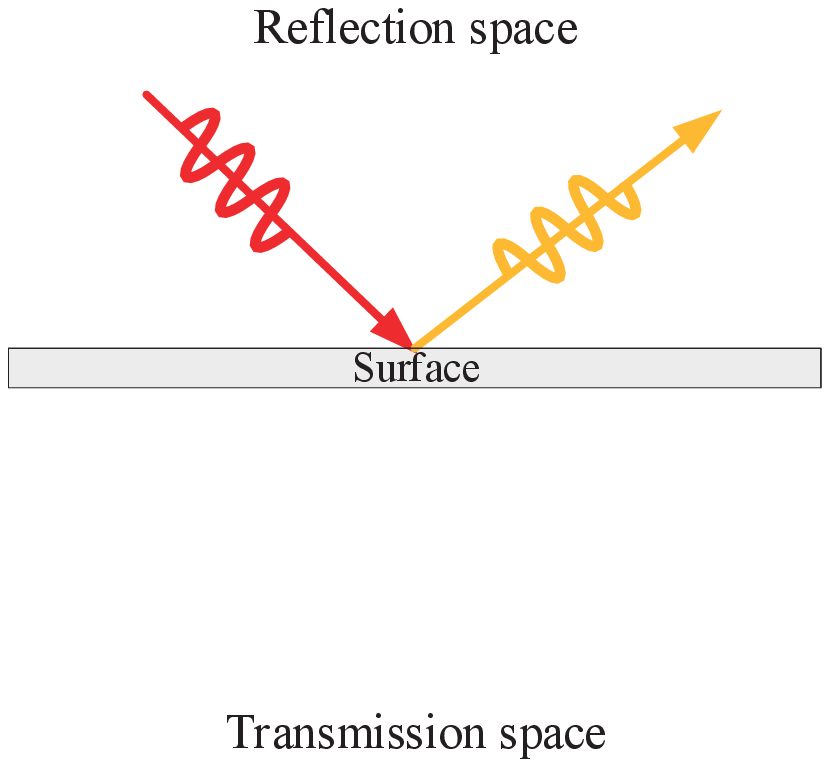}}
\subfigure[]{\label{FT}
\includegraphics[width= 1.1in]{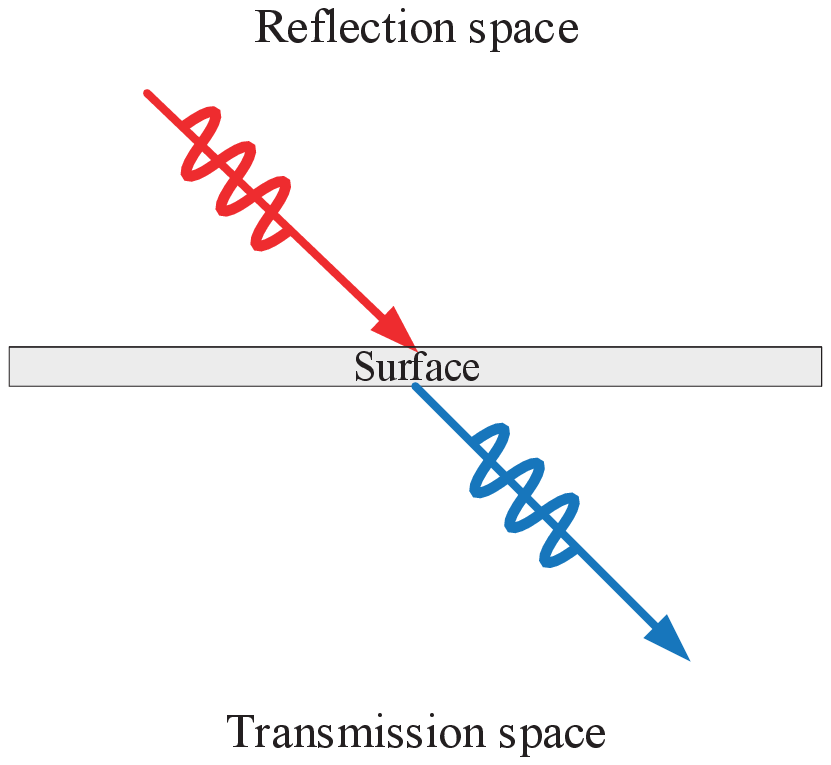}}
\subfigure[]{\label{STR}
\includegraphics[width= 1.1in]{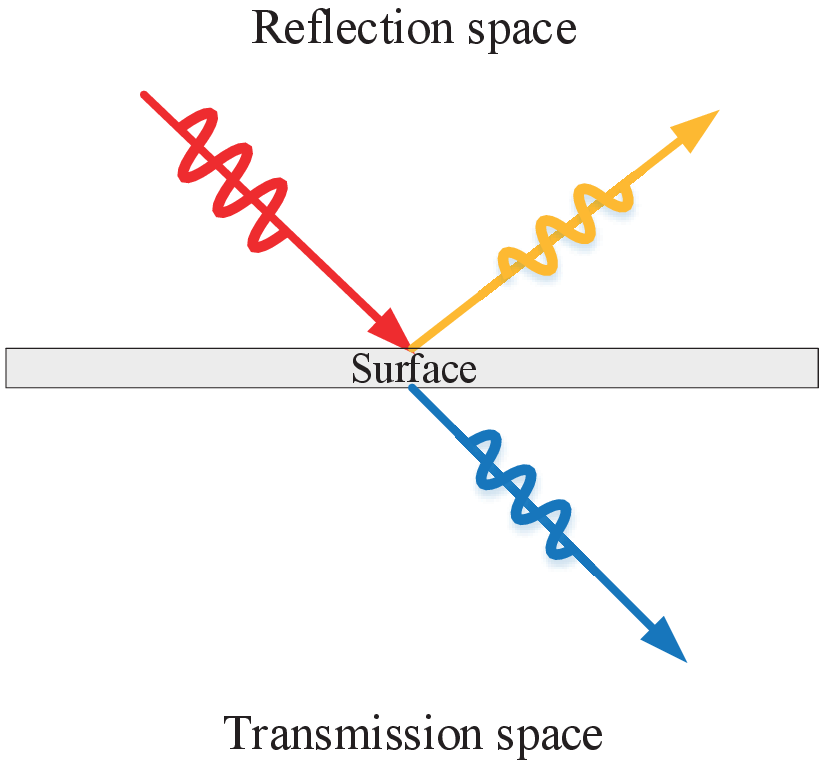}}
\caption{Illustration of signal propagation types of STAR-RISs. (a) NTT DOCOMO's prototype (photo: NTT DOCOMO) $\textcircled{c}$ CCBY~\cite{DOCOMO}, (b) full reflection (conventional reflecting-only RIS), (c) full transmission, (d) simultaneous transmission and reflection. }\label{propagation}
\end{figure*}
\subsection{From Conventional Reflecting-Only RISs to STAR-RISs: A Prototype}
We commence by providing a brief introduction to three types of signal propagation, namely full reflection, full transmission, as well as simultaneous transmission and reflection, based on a prototype developed by researchers from NTT DOCOMO, Japan~\cite{DOCOMO}. Fig. \ref{prototype} depicts NTT DOCOMO's prototype, where a metasurface is covered by a transparent substrate made of glass. By modifying the distance between the metasurface and the transparent substrate, the aforementioned three types of signal propagation can be achieved~\cite{DOCOMO}, as shown in Figs. \ref{FR}-\ref{STR}. For the full reflection scenario of Fig. \ref{FR}, the incident signals are completely reflected and cannot penetrate the surface. This type of wireless signal manipulation is widely investigated for conventional reflecting-only RISs. By contrast, for the full transmission scenario of Fig. \ref{FT}, all incident signals pass through the surface into the transmission space, while no signal is reflected. Finally, for the simultaneous transmission and reflection scenario of Fig. \ref{STR}, the incident signals are divided into two parts by the surface. Part of the signal is reflected to the reflection space, while the remaining part is radiated into the transmission space, thus facilitating the full-space manipulation of signal propagation.
\subsection{Key Advantages and Motivations for Employing STAR-RISs in Wireless Communication Systems}
Considering the above unique features, the employment of STAR-RISs has the following advantages in wireless communication systems: 1) Thanks to their capability of simultaneously transmitting and reflecting the incident signals, the coverage of STAR-RISs is extended to the entire space, thus serving both half-spaces using a single RIS, which is not possible for conventional reflecting-only RISs. 2) STAR-RISs provide enhanced degrees-of-freedom (DoFs) for signal propagation manipulation, which significantly increases the design flexibility in satisfying stringent communication requirements. 3) Since STAR-RISs are generally designed to be optically transparent~\cite{DOCOMO}, they are aesthetically pleasing and readily compatible with existing building structures, such as windows. Therefore, STAR-RISs will have no undesired aesthetic effect, which is of vital importance for practical implementations.\\
\indent As noted above, the joint manipulation of transmission and reflection is not a new idea, especially from the perspectives of the physics and metasurface technology. Apart from the above NTT DOCOMO prototype~\cite{DOCOMO}, the authors of~\cite{Wang,IOS} have also proposed concepts similar to `STAR'~\cite{SRAR}. In~\cite{Wang}, the authors reported frequency-selective refection and transmission of signals by using a dual-band bi-functional metasurface structure. The authors of~\cite{IOS} proposed an intelligent omni-surface (IOS), where the signals transmitted and reflected by an IOS element are adjusted via a common phase shift. For the STAR-RISs in this work, the transmitted and reflected signals can be \emph{simultaneously} reconfigured by each element via two generally independent coefficients, namely the transmission and reflection coefficients~\cite{SRAR}. This distinct characteristic facilitates the flexible design of STAR-RIS-aided wireless networks. However, the wireless communication design of STAR-RISs is still in its infancy. This motivates us to provide a systematic introduction to STAR-RISs, including the fundamental differences to conventional reflecting-only RISs, the basic signal model and practical operating protocols for STAR-RISs, complemented by their promising application scenarios and their performance evaluation.\\
\indent The main contributions of this article can be summarized as follows.
\begin{itemize}
  \item The differences between conventional reflecting-only RISs and STAR-RISs are identified from the perspectives of hardware design, physics principles, and communication system design. In particular, the importance of exploiting both the electric and magnetic properties for achieving `STAR' is highlighted.
  \item A basic signal model for incorporating STAR-RISs in wireless communications is introduced and three practical protocols for operating STAR-RISs, namely energy splitting (ES), mode switching (MS), and time switching (TS), are presented along with their benefits and drawbacks.
  \item A range of promising application scenarios of STAR-RISs in wireless networks are proposed both in outdoor and indoor environments.
  \item The performance of STAR-RISs for different operating protocols is studied and compared with those of other baselines in a downlink multiple-input single-output (MISO) system for both unicast and multicast scenarios.
\end{itemize}
\begin{figure}[!t]
\centering
\subfigure[Reflecting-only RISs (biscuits placed on a metal plate)]{\label{2a}
\includegraphics[width= 3in]{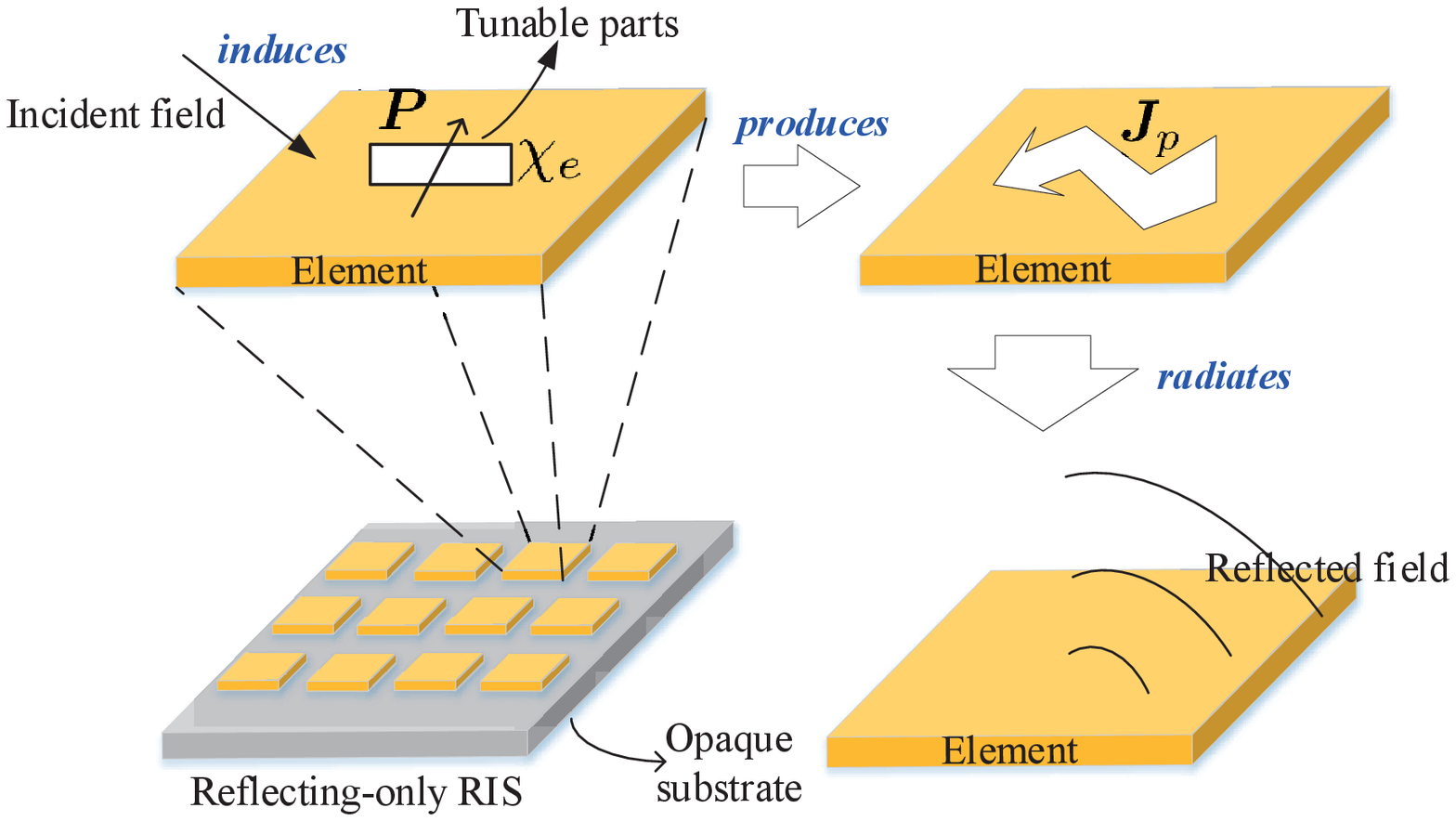}}
\subfigure[STAR-RISs (ice cubes in a glass of water)]{\label{2b}
\includegraphics[width= 3in]{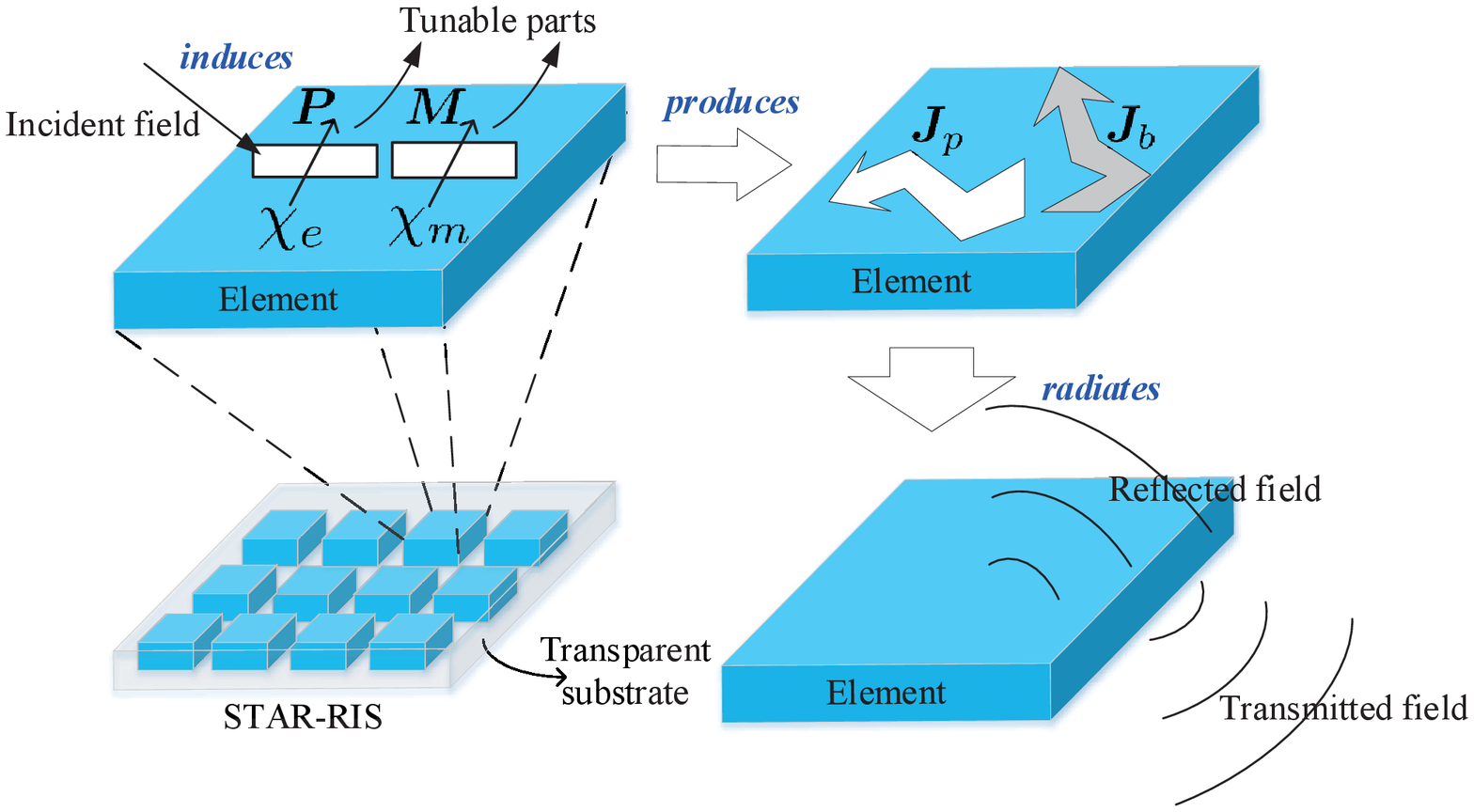}}
\caption{Conceptual comparison between reflecting-only RISs and STAR-RISs.}\label{fig_new}
\end{figure} 
\section{Key Differences between Reflecting-Only RISs and STAR-RISs}
In this section, we discuss the key differences between conventional reflecting-only RISs and the proposed STAR-RISs from their hardware design, physics principles, and communication system design perspectives, respectively. We highlight that STAR-RISs rely on substrates, which are transparent at radio frequency and have elements, which support magnetic currents. These structural properties allow STAR-RISs to achieve simultaneous and independent control of their transmission and reflection coefficients.
\subsection{Hardware Design Differences}
Reflecting-only RISs and STAR-RISs are different both in terms of their equipped elements and substrates. The following analogy illustrates the structural differences between reflecting-only RISs and STAR-RISs. For reflecting-only RISs, the reconfigurable elements on the substrate are like \textbf{\textit{biscuits placed on a metal plate}}, as illustrated in Fig.~\ref{2a}, while, for STAR-RISs, the reconfigurable elements are like \textbf{\textit{ice cubes in a glass of water}}, as illustrated in Fig.~\ref{2b}. To elaborate, the substrates of reflecting-only RISs are opaque for wireless signals at their operating frequency. The opaque substrate serves as a bed, on which the tunable elements are integrated. It also prevents the wireless signals from penetrating the RIS so that no energy is leaked into the space behind the RIS. By contrast, the substrates of STAR-RISs have to be transparent for wireless signals at their operating frequency. Naturally, upon facilitating simultaneous transmission and reflection, STAR-RISs require a more complex design, since their elements have to support both electric and magnetic currents. Consequently, these elements have to be thicker than those of conventional reflecting-only RISs~\cite{monticone2013full}.
\subsection{Physics Principles Differences}\label{principle}
Again, compared to conventional reflecting-only RISs, STAR-RISs must have elements, which support both electric polarization currents $\bm{J}_p$ and magnetization currents $\bm{J}_b$~\cite{pfeiffer2013metamaterial,hao1}. The physical principles behind STAR-RISs can be summarized in three steps, namely \textit{\textbf{induction}}, \textit{\textbf{production}}, and \textit{\textbf{radiation}}, as illustrated in Fig.~\ref{fig_new}.
\begin{itemize}
\item Firstly, the elements are polarized by the incident field. The patch-elements of pure reflecting RISs only respond to the electric component of the incident field and a polarization density $\bm{P}$ is induced. By contrast, the elements of STAR-RISs respond to both the electric and magnetic components of the incident field. Hence, both a polarization density $\bm{P}$ and a magnetization density $\bm{M}$ are \textit{\textbf{induced}}. The strengths of the polarization and magnetization densities depend on the electric susceptibility $\chi_e$ and magnetic susceptibility $\chi_m$, respectively. The tunable parameters of the elements can be used to adjust the values of these susceptibilities within a certain quantization error.
\item Secondly, the oscillating polarization and magnetization densities \textit{\textbf{produce}} time-varying electric polarization and magnetization currents on the surface, respectively.
\item Lastly, these time-varying currents \textit{\textbf{radiate}} both the transmitted and reflected fields back into free-space, producing phase differences between the incident field and the transmitted or reflected fields.
\end{itemize}
\indent As illustrated in Fig.~\ref{2a}, reflecting-only RISs having non-magnetic elements can only support surface electric polarization currents. If the elements consist of only single-layered metallic scatters (not considering the substrate), the radiated fields on both sides of the RIS are identical~\cite{mohammadi2015graded}. Thus, this symmetry limitation of non-magnetic RISs does not facilitate the independent control of the transmitted and reflected signals. On the contrary, by also supporting magnetic currents, STAR-RISs break this symmetry limitation and can achieve simultaneous control of both the transmitted and reflected signals~\cite{hao1}. As illustrated in Fig.~\ref{2b}, assuming the electric and magnetic susceptibilities of each element are constant, the magnetization density $\bm{M}$ introduces extra DoFs by enabling the independent adjustment of the phase shift for transmission.
\subsection{Communication System Design Differences}
From the perspective of communication system design, the benefits of supporting surface magnetic currents in STAR-RISs can be exploited as follows.
\begin{itemize}
\item \textit{\textbf{Adjustable energy ratio}}: The amplitudes of the transmitted and reflected waves of each STAR-RIS element can be dynamically adjusted, respectively. Since the overall energy of the transmitted and reflected signals are the magnitudes of the complex-valued sum of the contributions of all elements, the energy ratio between the transmitted and reflected signals can be controlled.
\item \textit{\textbf{Independent beamforming}}: Again, the introduction of surface magnetic currents enables independent phase shift control for transmitted and reflected signals. As a result, STAR-RISs allow independent transmission and reflection beamforming for the two half-spaces, thus improving the flexibility of communication system design.
\end{itemize}
\section{Basic Signal Model and Practical Operating Protocols}
In this section, we introduce the basic signal model for STAR-RISs, and then propose three practical operating protocols for integrating STAR-RISs into wireless communication systems, while identifying their respective advantages and disadvantages.
\subsection{Basic Signal Model}
\begin{figure*}[!t]
\centering
\subfigure[Energy splitting (ES).]{\label{ES}
\includegraphics[width= 1.9in, height= 1.5in]{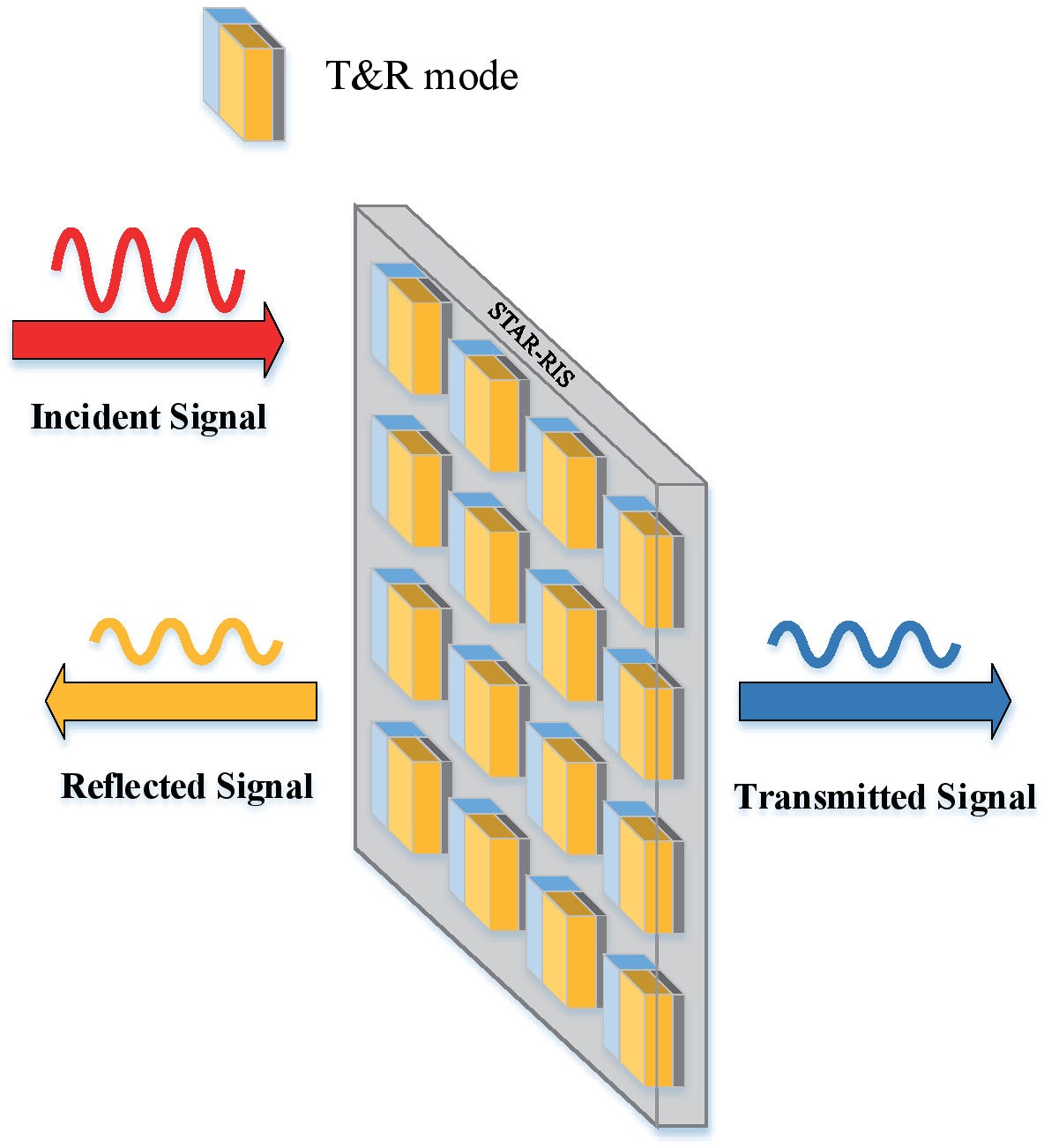}}
\subfigure[Mode switching (MS).]{\label{MS}
\includegraphics[width= 1.9in, height= 1.5in]{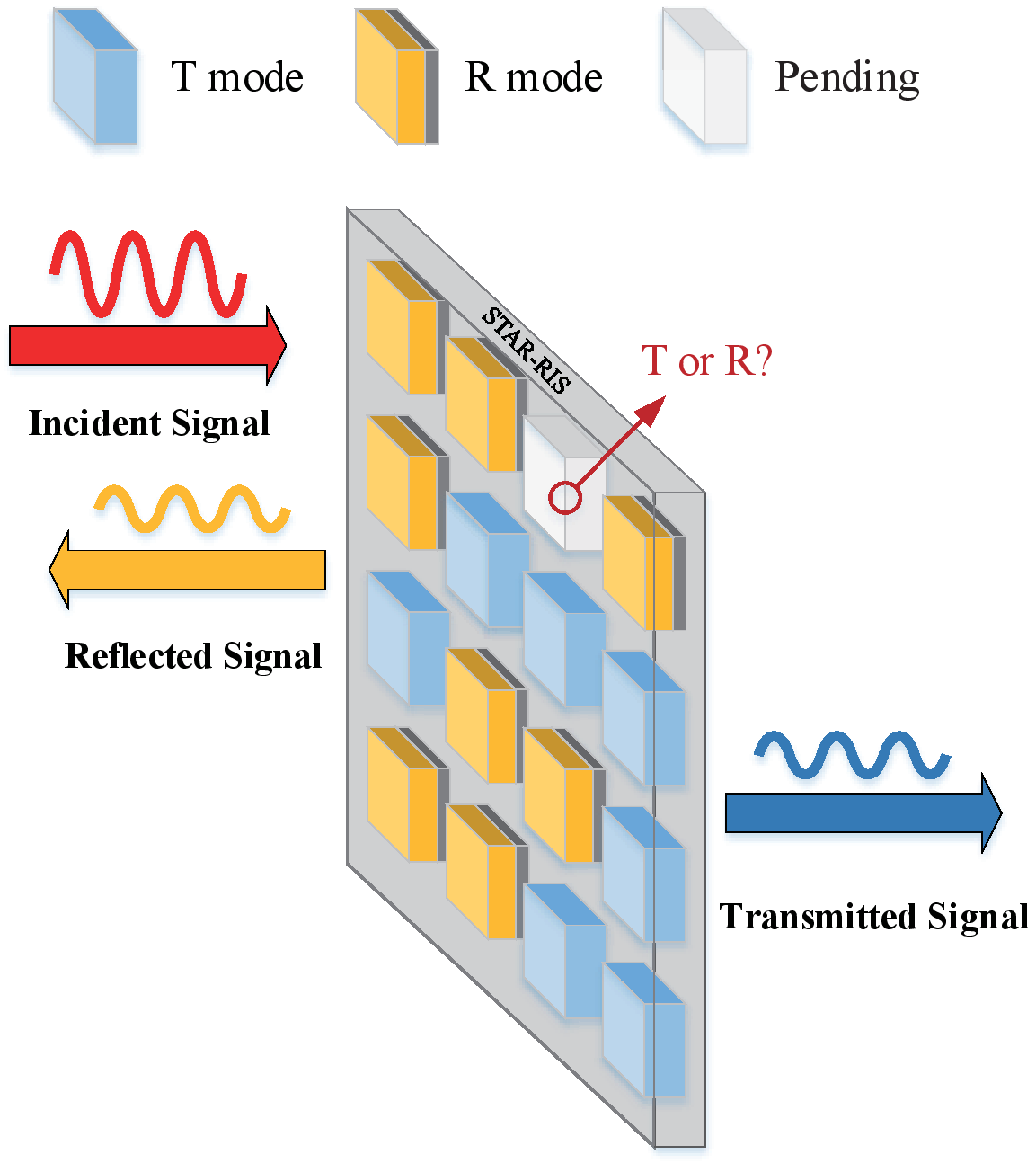}}
\subfigure[Time switching (TS).]{\label{TS}
\includegraphics[width= 1.9in, height= 1.5in]{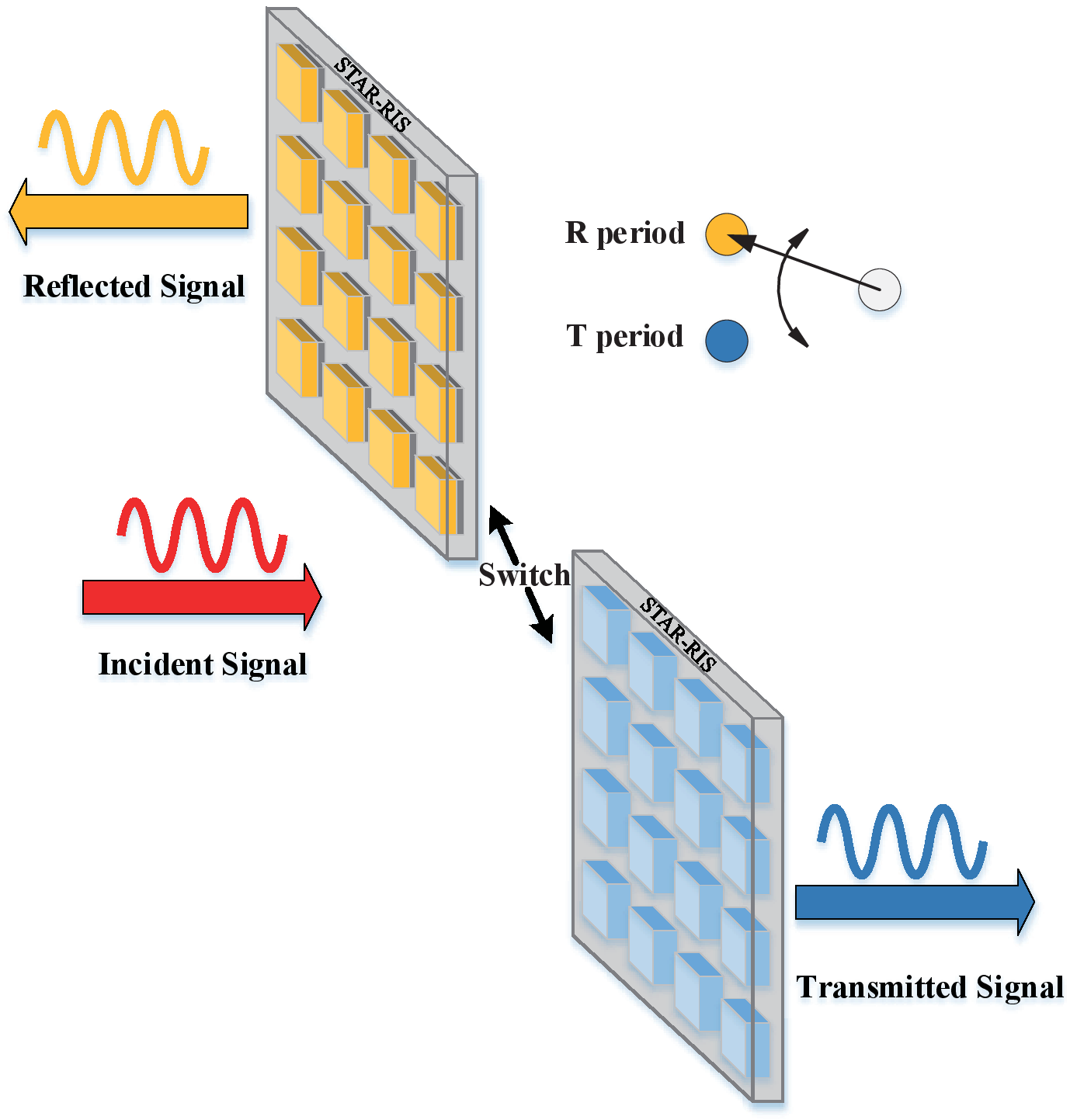}}
\caption{Illustration of three practical protocols for operating STAR-RISs.}\label{structure}
\end{figure*}

As discussed in the previous section, STAR-RISs are capable of independently controlling the transmitted and reflected signals, which introduces additional DoFs that can be exploited. To characterize this unique feature, for a STAR-RIS having $M$ elements, let $s_m$ denote the signal incident upon the $m$th element. After being reconfigured by the corresponding transmission and reflection coefficients, the signals transmitted and reflected by the $m$th element are given by $\left( {\sqrt {\beta _m^t} {e^{j\theta _m^t}}} \right){s_m}$ and $\left( {\sqrt {\beta _m^r} {e^{j\theta _m^r}}} \right){s_m}$, respectively. In particular, $\sqrt {\beta _m^t} ,\sqrt {\beta _m^r}  \in \left[ {0,1} \right]$ and $\theta _m^t,\theta _m^r \in \left[ {0,2\pi } \right)$ characterize the amplitude and phase shift adjustments imposed on the incident signal facilitated by the $m$th element during transmission and reflection, respectively. The adjustments of the phase shifts for transmission and reflection can be generally chosen independent of each other~\cite{SRAR}. However, the transmission and reflection amplitude coefficients must obey the law of energy conservation, i.e., the sum of the energies of the transmitted and reflected signals has to be equal to the incident signal's energy, which leads to the sum of $\beta _m^t$ and $\beta _m^r$ should be equal to one~\cite{SRAR}. Accordingly, by adjusting the amplitude coefficients for transmission and reflection, each STAR-RIS element can be operated in full transmission mode (referred to as T mode), full reflection mode (referred to as R mode), and simultaneous transmission and reflection mode (referred to as T\&R mode). Based on the above basic signal model of STAR-RISs, in the following, we propose three practical protocols for operating STAR-RISs in wireless networks, namely energy splitting (ES), mode switching (MS), and time switching (TS), as illustrated in Fig. \ref{structure}.
\begin{table*}[!t]
\caption{Summary of Proposed Protocols for Operating STAR-RISs}
\begin{center}
\centering
\resizebox{\textwidth}{!}{
\begin{tabular}{|l|l|l|l|}
\hline
\centering
\textbf{Protocols}  & \textbf{Optimization Variables} &\textbf{Advantages} & \textbf{Disadvantages} \\
\hline
\centering
ES & \makecell[l]{$\bullet$ Amplitude and phase shift coefficients of each element\\ $\;\;$ for transmission and reflection}  & High flexibility & Large number of design variables  \\
\hline
\centering
MS&  \makecell[l]{$\bullet$ Mode selection of each element \\ $\bullet$ Transmission phase shift coefficients for T mode elements\\ $\bullet$ Reflection phase shift coefficients for R mode elements}  & Easy to implement & \makecell[l]{Reduced transmission and reflection gain} \\
\hline
\centering
TS&  \makecell[l]{$\bullet$ Time allocation \\ $\bullet$ Transmission phase shift coefficients of each element during T period \\ $\bullet$ Reflection phase shift coefficients of each element during R period}  & \makecell[l]{Independent T and R design}    & High hardware implementation complexity  \\
\hline
\end{tabular}
}
\end{center}
\label{table:structure}
\end{table*}
\subsection{Energy Splitting}
For ES, all elements of the STAR-RIS are assumed to operate in T\&R mode, as shown in Fig. \ref{ES}. For given transmission and reflection amplitude coefficients, the signals incident upon each element are split into transmitted and reflected signals having different energy. In a practical implementation, the amplitude and phase shift coefficients of each element for transmission and reflection can be jointly optimized for achieving diverse design objectives in wireless networks.
\subsection{Mode Switching}
In MS, all elements of the STAR-RIS are partitioned into two groups. Specifically, one group contains the elements that operate in T mode, while the other group contains the elements operating in R mode. As shown in Fig. \ref{MS}, an MS STAR-RIS can be viewed as being composed of a conventional reflecting-only RIS and a transmitting-only RIS of reduced sizes. For this protocol, the element-wise mode selection and the corresponding transmission and reflection phase shift coefficients can be jointly optimized. The resulting ``on-off'' type of protocol (i.e., transmission or reflection) makes MS easy to implement. However, the drawback is that MS generally cannot match the transmission and reflection gain of ES, since only a subset of the elements are selected for transmission and reflection, respectively.
\subsection{Time Switching}
By contrast, the STAR-RIS employing the TS protocol periodically switches all elements between the T mode and R mode in orthogonal time slots (referred to as T period and R period), as illustrated in Fig. \ref{TS}. This is like switching ``venetian blinds'' in different time slots. The fraction of time allocated to fully transmitting and fully reflecting signals can be optimized to strike a balance between the communication qualities of the front and back sides. Compared to ES and MS, the advantage of TS is that, for a given time allocation, the transmission and reflection coefficients are not coupled, hence they can be optimized independently. Nevertheless, periodically switching the elements imposes stringent time synchronization requirements, thus increasing the implementation complexity compared to the ES and MS.\\
\indent In Table \ref{table:structure}, we summarize the unique ES, MS, and TS optimization variables, and identify their respective advantages and disadvantages.
\section{Promising Applications of STAR-RISs in 6G}
Having presented practical protocols for operating STAR-RISs, in this section, we discuss several attractive applications of STAR-RISs in next-generation networks for both outdoor and indoor environments, as illustrated in Fig. \ref{application}.
\subsection{Outdoor, Outdoor-Indoor, and Indoor Coverage Extension}
\begin{figure*}[!t]
  \centering
  \includegraphics[width= 5in]{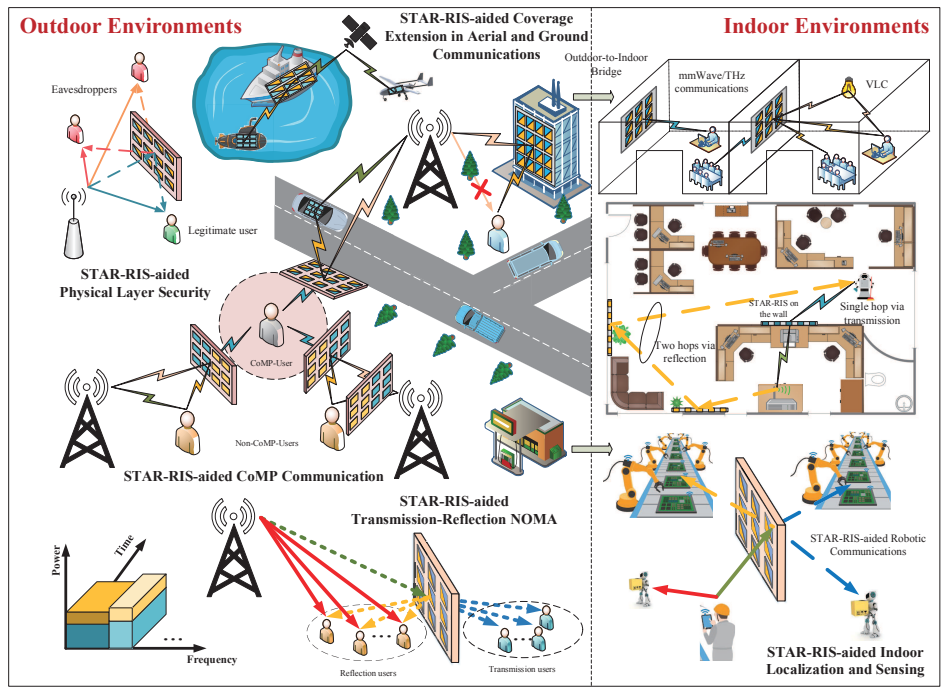}\\
  \caption{Illustration of application scenarios of STAR-RISs in wireless communications for outdoor and indoor environments.}\label{application}
\end{figure*}

One of the most promising applications of STAR-RISs is to improve the coverage area/quality of wireless networks, especially when the links between the base stations (BSs) or access points (APs) and users are severely blocked by obstacles (e.g., trees along roads, buildings, and metallic shells of vehicles). As shown at the top right of Fig. \ref{application}, STAR-RIS-aided coverage extension can be loosely divided into three scenarios, namely outdoor, outdoor-to-indoor, and indoor.\\
\indent In outdoor communications, similar to conventional reflecting-only RISs, STAR-RISs can be mounted on building facades and roadside billboards to create an additional communication link. More innovatively, STAR-RISs can also be accommodated by the windows of vehicles (e.g., cars, aircraft, and cruise ships) to enhance the signal strength received inside by exploiting their transmission capability, thus extending the coverage area/quality of BSs and satellites. For outdoor-to-indoor communications, the severe penetration loss caused by building walls gravely restricts the coverage provided by outdoor BSs, especially in mmWave and THz communications. In fact, STAR-RISs constitute an efficient technique for creating an outdoor-to-indoor \emph{bridge} as illustrated in Fig. \ref{application}. For indoor communications, STAR-RISs are more appealing than conventional reflecting-only RISs. As conventional reflecting-only RISs merely achieve half-space coverage, the signals emerging from the AP may require multi-hop bounces for reaching the target user. However, by exploiting both transmission and reflection, the resultant full-space coverage may reduce the propagation distance, thus increasing the received signal power. An example, where conventional reflecting-only RISs require two hops, whereas the STAR-RIS needs only a single hop, is illustrated at middle right of Fig. \ref{application}. In a nut shell, STAR-RISs substantially outperform conventional reflecting-only RISs, since they do not only possess the same capabilities as conventional reflecting-only RISs but also support additional design options due to their transmission capability.
\subsection{Transmission-Reflection NOMA}
Non-orthogonal multiple access (NOMA) is a promising next-generation candidate facilitating flexible resource allocation, high spectrum efficiency, and supporting massive connectivity. For NOMA to achieve a large performance gain over orthogonal multiple access (OMA), it is important to pair users having different channel conditions. However, for conventional reflecting-only RISs, the benefits of NOMA may not be fully reaped since the channel conditions of users in the local reflected space are generally similar. Exploiting STAR-RISs enables a novel communication framework, namely transmission-reflection NOMA, where a pair of users at the transmission- and reflection-oriented side can be grouped together for facilitating NOMA, as shown at the bottom left of Fig. \ref{application}. By carefully optimizing the element-based energy splitting ratio of the proposed ES protocol or the element-based mode selection of the proposed MS protocol, sufficiently different transmitted and reflected channel conditions can be achieved. As a result, the proposed STAR-RIS-aided transmission-reflection NOMA framework is capable of achieving higher gain over conventional reflecting-only RIS-aided NOMA.
\subsection{CoMP Communication via Transmission and Reflection}
For realistic multi-cell communication networks, the performance of cell-edge users cannot be guaranteed due to the strong inter-cell interference. Coordinated multi-point (CoMP) communication efficiently mitigates the inter-cell interference. In the middle left of Fig. \ref{application}, a beneficial STAR-RIS-aided CoMP scenario is presented. In particular, several multiple-antenna BSs are employed for cooperatively supporting a cell-edge user, referred to as CoMP-user. Additionally, each BS serves an additional cell-center user, referred to as non-CoMP-user. A STAR-RIS is deployed in each cell where the CoMP-user is located in the transmission half-space while the non-CoMP-user is located in the reflection half-space. The advantages of this implementation are that, on the one hand, the received signal-to-interference-plus-noise ratio (SINR) of the CoMP-user can be enhanced through the design of the cooperative transmission coefficients of all STAR-RISs, while on the other hand, the reflection coefficients of each STAR-RIS can be optimized for mitigating the intra-cell interference received from the cell-edge user at each non-CoMP-user.
\subsection{Full-space Physical Layer Security}
RISs are also capable of improving the physical layer security (PLS), where the channel conditions of the eavesdroppers can be degraded by degrading their signal propagation. However, for conventional reflecting-only RISs aided secure communication, the legitimate users and eavesdroppers are assumed to be located at the same side of the RISs, even though this idealized simplifying assumption may not hold in practice. Fortunately, STAR-RISs come to the rescue. Observe at the top left of Fig. \ref{application}, with the assistance of full-space STAR-RIS propagation, PLS can be enhanced, regardless of the eavesdropper location.
\subsection{Indoor Localization and Sensing}
By overcoming signal blockages and providing full-space coverage, STAR-RISs are capable of improving both the localization and sensing capability of wireless networks, especially in indoor environments. As illustrated at the bottom right of Fig. \ref{application}, the employment of STAR-RISs in smart factories improves the positioning of mobile robots and the data-rate of control links.\\
\indent There are also other promising application scenarios for STAR-RISs in 6G networks, such as STAR-RIS-aided simultaneous wireless information and power transfer (SWIPT), STAR-RIS-assisted visible light communications (VLCs), STAR-RIS-aided mmWave/THz communications, and STAR-RIS-augmented robotic communications. These applications constitute interesting future research directions.
\section{Numerical Case Studies}
\begin{figure*}[!t]
\centering
\subfigure[Simulation setup.]{\label{setup}
\includegraphics[width= 2in, height= 1.4in]{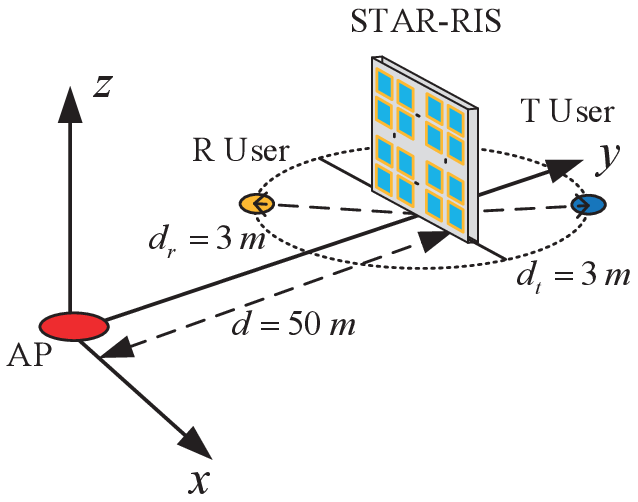}}
\subfigure[Unicast communication.]{\label{unicast}
\includegraphics[width= 2in, height= 1.4in]{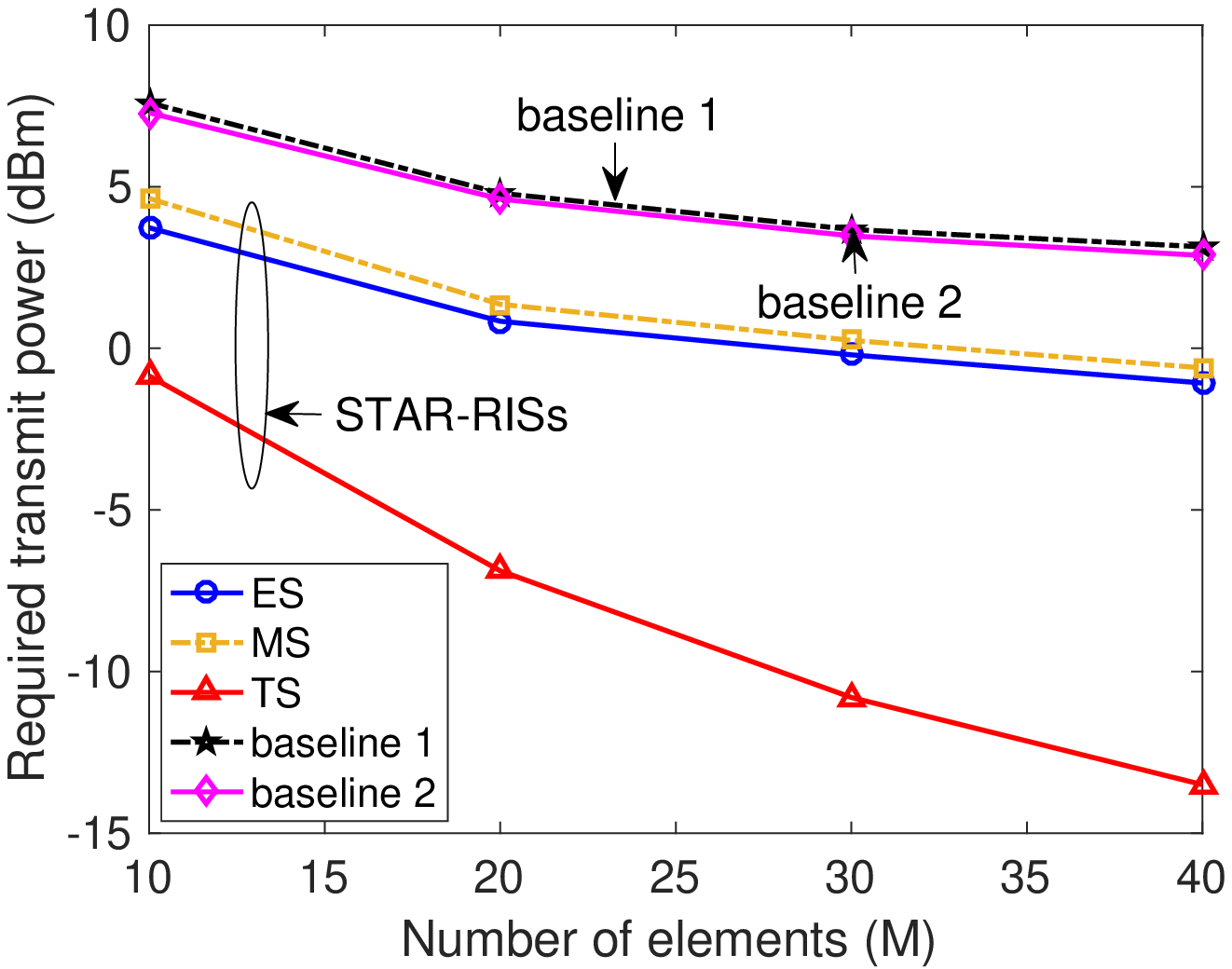}}
\subfigure[Multicast communication.]{\label{multicast}
\includegraphics[width= 2in, height= 1.4in]{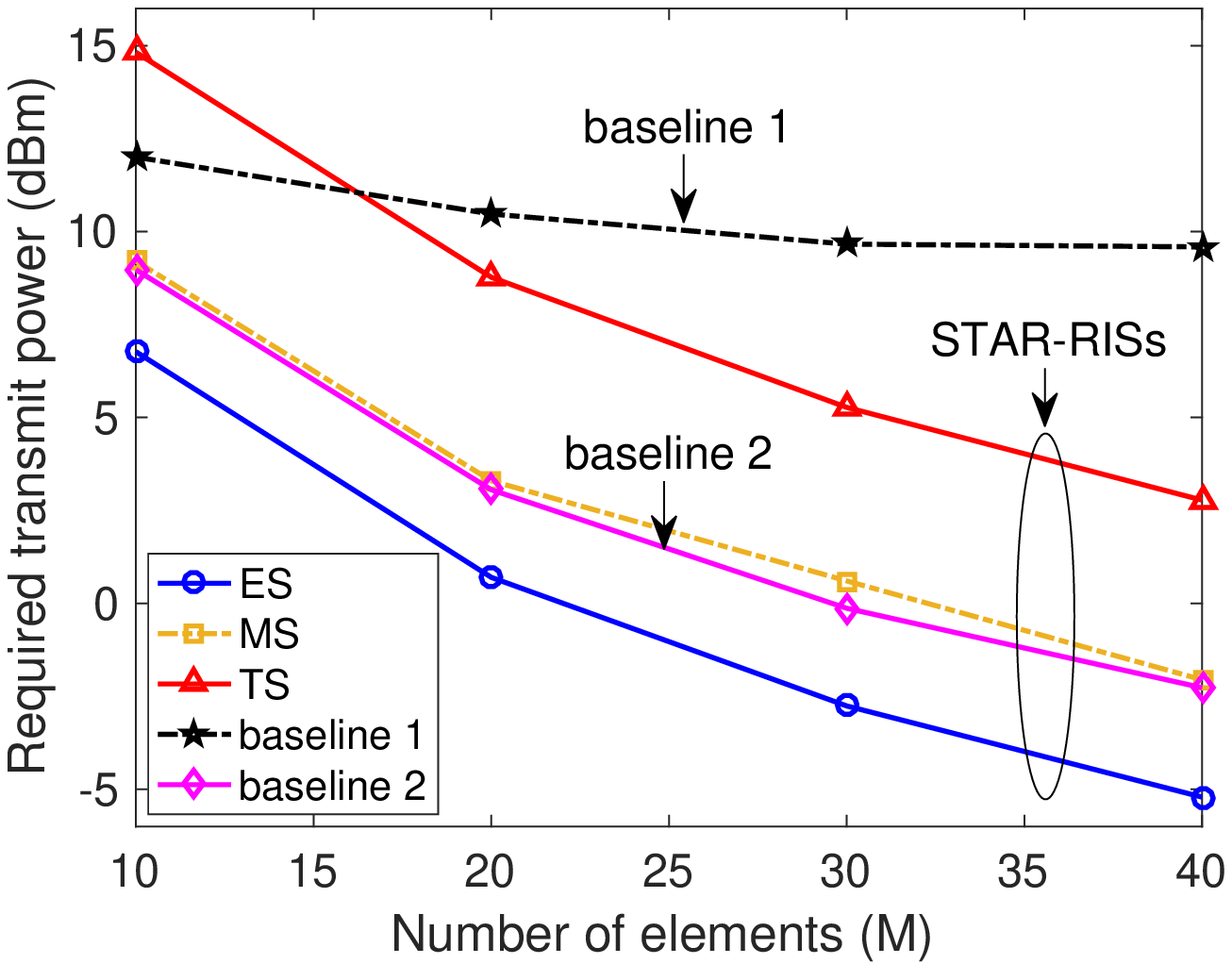}}
\caption{Transmit power versus the number of elements in a STAR-RIS-aided downlink MISO network. The target rates of the users are set to 1 bit/s/Hz and 3.46 bit/s/Hz in the unicast and multicast scenarios, respectively.}\label{transmit power}
\end{figure*}

In this section, we present numerical examples to characterize the performance of STAR-RISs employing the proposed operating protocols and to compare them with other baselines. More specifically, we consider the case where a two-antenna AP communicates with two single-antenna users with the aid of a STAR-RIS having $M$ elements. One of the user is assumed to be located in the STAR-RIS's transmission half-space, referred to as T user, and the other user is assumed to be located in the reflection half-space, termed as R user. The geographical setup is shown in Fig. \ref{setup}. The direct AP-user links are assumed to be blocked and only the STAR-RIS transmission/reflection-side AP-user links are available, which are assumed to obey the Rician fading channel model having a path-loss exponent of 2.2.\\
\indent For the considered setup, we investigate both unicast and multicast scenarios. In particular, the AP sends different messages to different users in the unicast scenario, while conveying a common message to both users in the multicast scenario. The minimum transmit power required by the AP for satisfying target user rates versus the number of STAR-RIS elements is studied. For comparison, two baselines are considered. For baseline 1, the full-space coverage is achieved by employing one conventional reflecting-only RIS and one transmitting-only RIS, each of which has ${M \mathord{\left/
 {\vphantom {M 2}} \right.
 \kern-\nulldelimiterspace} 2}$ elements. For baseline 2, the full-space coverage is achieved by an omni-surface proposed in \cite{IOS}, where each element is assumed to have identical transmission and reflection coefficients.\\
\indent Fig. \ref{unicast} and Fig. \ref{multicast} show that, for STAR-RISs with different operating protocols and those baseline schemes in both scenarios, the minimum transmit power required by the AP decreases upon increasing $M$, since a higher transmission/reflection gain can be achieved. Regarding the performance of STAR-RISs, it is interesting to observe that TS achieves the best performance in the unicast scenario, whereas ES is preferable in the multicast scenario. This is because TS achieves interference-free communication for each user in the unicast scenario. By contrast, since the multicast scenario does not introduce inter-user interference, ES exploits the entire available communication time while TS cannot. It is also observed that ES always outperforms MS. This is indeed expected, since from a theoretical point of view, MS can be regarded as a special case of ES with binary amplitude coefficients for each element. These results highlight the importance of employing different operating protocols for satisfying different communication objectives.\\
\indent Furthermore, regarding the performance comparison, it can be observed that, independent of the adopted operating protocols, STAR-RISs always outperform conventional RISs and the omni-surface in the unicast scenario of Fig. \ref{unicast}. Since conventional RISs employ fixed element-based mode selection and each omni-surface element employs identical transmission and reflection coefficients, the two schemes cannot fully exploit the DoFs available at each element to enhance the desired signal strength and mitigate the inter-user interference as STAR-RISs can, and thus yielding the worst performance for unicast. For the multicast scenario of Fig. \ref{multicast}, conventional RISs only outperform TS STAR-RIS for small $M$ and the omni-surface achieves a similar performance with MS STAR-RIS. As the conventional RISs setup is a special case of MS STAR-RISs, unlike TS, it can fully exploit the available communication time. This advantage allows conventional RISs to achieve a higher performance than TS when $M$ is small, i.e., the case where the available DoFs are limited for both schemes and using the entire available communication time dominates the achieved performance. However, when $M$ increases, it is expected that conventional RISs becomes the worst option again due to the significant loss of DoFs. This drawback also causes the performance gap between conventional RISs and STAR-RISs to become more pronounced as $M$ increases. For the omni-surface, due to the absence of inter-user interference in multicast, the performance degradation caused by employing common transmission and reflection coefficients alleviates. The fully exploitation of communication time and the full-dimension transmission/reflection gain respectively enables the omni-surface to significantly outperform TS STAR-RIS and to slightly outperform MS STAR-RIS. However, omni-surface still achieves a worse performance than ES STAR-RIS due to the loss of DoFs. The provided performance comparisons confirm the effectiveness of employing the proposed STAR-RISs in wireless networks.
\section{Concluding Remarks and Future Research}
The unique differences between the proposed STAR-RISs and conventional reflecting-only RISs have been discussed from the perspectives of the hardware design, physics principles, and communication system design. Furthermore, three practical operating protocols have been proposed for STAR-RISs and their benefits and drawbacks have been highlighted. Moreover, several promising application scenarios for STAR-RISs in wireless networks have been identified both for outdoor and indoor environments. Through numerical case studies, the performance of STAR-RISs has been evaluated and compared with other baseline schemes in both unicast and multicast scenarios. However, the investigation of STAR-RISs is still at a very early stage, and there are numerous open research problems, some of which are exemplified below:
\begin{itemize}
  \item \textbf{Spatial Analysis of STAR-RIS Aided Networks using Stochastic Geometry}: The proposed STAR concept introduces new research challenges for the spatial analysis of RIS-aided large-scale networks, since transmission and reflection are determined by both the spatial locations and orientations of the involved nodes. To facilitate the application of the powerful stochastic geometry tools for performance analysis, new point processes that can capture the randomness of the angle-related spatial distributions have to be developed for the APs, STAR-RISs, and users.
  \item \textbf{Channel Estimation for STAR-RISs}: Due to the near-passive nature of RISs, the acquisition of accurate channel state information is a non-trivial task in the face of having both transmission and reflection links. On the one hand, based on the TS protocol, the transmission and reflection links can be consecutively estimated, thus achieving high accuracy, but at the cost of a considerable pilot-overhead. On the other hand, the channels for the two links can also be simultaneously estimated based on the ES protocol, hence reducing the pilot-overhead. The development of efficient channel estimation methods for striking an attractive tradeoff between the performance and overhead requires further research.
  \item \textbf{Deployment Strategies for STAR-RISs}: Since STAR-RISs provide full-space 360$^{\circ}$ coverage, the corresponding deployment design problem raises a lot of challenging questions, especially for realistic multi-user scenarios. The deployment locations of STAR-RISs have to be carefully chosen to balance the number of users in the transmission and reflection half-spaces. This constitutes an interesting but challenging new research problem.
\end{itemize}

\bibliographystyle{IEEEtran}

\begin{thebibliography}{1}

\bibitem{Renzo2020JSAC}
M.~{Di~Renzo}, \emph{et al.}, ``Smart radio environments empowered by reconfigurable intelligent surfaces: How it works, state of research, and road ahead,'' \emph{IEEE J. Sel. Areas Commun}., vol. 38, no. 11, pp. 2450--2525, 2020.

\bibitem{di2019smart}
M.~{Di~Renzo}, \emph{et al.}, ``Smart radio environments empowered by {AI} reconfigurable meta-surfaces: An
  idea whose time has come,'' \emph{EURASIP J. Wireless Commun.}, 2019.

\bibitem{Huang_Holographic}
C.~{Huang}, \emph{et al.}, ``Holographic {MIMO} surfaces for {6G} wireless networks: Opportunities, challenges, and trends,'' \emph{IEEE Wireless Commun.}, vol.~27, no.~5, pp.~118--125, 2020.

\bibitem{Wu_MAG}
Q.~{Wu} and R.~{Zhang}, ``Towards smart and reconfigurable environment: Intelligent reflecting surface aided wireless network,'' \emph{{IEEE} Commun. Mag.}, vol.~58, no.~1, pp. 106--112, 2020.

\bibitem{Huang_EE}
C.~{Huang}, \emph{et al.}, ``Reconfigurable intelligent surfaces for energy efficiency in wireless communication,'' \emph{{IEEE} Trans. Wireless Commun.}, vol.~18, no.~8, pp.~4157--4170, 2019.

\bibitem{Akyildiz}
I.~F.~{Akyildiz}, C.~{Han}, and S.~{Nie}, ``Combating the distance problem in the millimeter wave and terahertz frequency bands,'' \emph{IEEE Commun. Mag.}, vol.~56, no.~6, pp.~102--108, 2018.

\bibitem{Huang_AI}
C.~{Huang}, R.~{Mo}, and C.~{Yuen}, ``Reconfigurable intelligent surface assisted multiuser {MISO} systems exploiting deep
reinforcement learning,'' \emph{IEEE J. Sel. Areas Commun.}, vol.~38, no.~8, pp.~1839--1850, 2020.

\bibitem{SRAR}
J.~Xu, \emph{et al.}, ``SRAR-RISs: Simultaneous reflecting and refracting reconfigurable intelligent surfaces,'' Feb. 2021. [Online]. Available: \url{https://arxiv.org/abs/2101.09663}.

\bibitem{DOCOMO}
NTT DOCOMO, ``DOCOMO conducts world's first successful trial of transparent dynamic metasurface,'' Jan. 2020. [Online]. Available:
\url{https://www.nttdocomo.co.jp/english/info/media\_center/pr/2020/0117\_00.html}.

\bibitem{Wang}
X.~{Wang}, \emph{et al.}, ``Simultaneous realization of anomalous reflection and transmission at two frequencies using bifunctional metasurfaces.'' \emph{Sci. Rep.}, vol.~8, no.~1, pp.~1--8, 2018.

\bibitem{IOS}
S. Zhang, \emph{et al.}, ``Beyond intelligent reflecting surfaces: Reflective-transmissive metasurface aided communications for full-dimensional coverage extension'', \emph{IEEE Trans. Veh. Technol.}, vol.~69, no.~11, pp.~13905--13909, 2020.

\bibitem{monticone2013full}
F.~Monticone, N.~M.~Estakhri, and A.~Al\`{u}, ``Full control of nanoscale optical transmission with a composite metascreen,'' \emph{Phys. Rev. Lett.}, vol.~110, no.~20, pp.~203903, 2014.

\bibitem{pfeiffer2013metamaterial}
C.~{Pfeiffer} and A.~{Grbic}, ``Metamaterial Huygens' surfaces: Tailoring wave fronts with reflectionless sheets,'' \emph{Phys. Rev. Lett.}, vol.~110, no.~19, pp.~197401, 2013.

\bibitem{hao1}
L.~{La~Spada}, \emph{et al.}, ``Curvilinear metasurfaces for surface wave manipulation,'' \emph{Sci. Rep.}, vol.~9, no.~1, pp.~1--10, 2019.

\bibitem{mohammadi2015graded}
N.~M.~{Estakhri}, C.~{Argyropoulos}, and A.~Al\`{u}, ``Graded metascreens to enable a new degree of nanoscale light management'' \emph{Phil. Trans. R. Soc. A.}, vol.~373, no.~2049, pp.~20140351, 2015.

\end{thebibliography}

\end{document}